  \documentclass[]{article}
\usepackage{graphicx} 
\usepackage{subfigure} 
\usepackage{lipsum}

\usepackage{natbib}
\usepackage{algpseudocode}
\usepackage{algorithm}
\usepackage{enumitem}
\usepackage{color}
\usepackage{url}
\usepackage{amsmath}
\usepackage{amssymb}
\usepackage{xspace}
\usepackage{amsfonts}
\usepackage{booktabs}
\usepackage{siunitx}
\usepackage{tabularx}
\usepackage[T1]{fontenc}
\usepackage[utf8]{inputenc}
\usepackage{graphicx}
\usepackage[flushleft]{threeparttable}

\usepackage{array}
\usepackage{calc}

\newcommand{\x}[1]{$x_{#1}$\xspace}

\newcommand{\fe}{\textit{feature}\xspace}
\newcommand{\fes}{\textit{features}\xspace}
\newcommand{\cov}{covariates\xspace}
\newcommand{\coeff}{coefficients\xspace}
\newcommand{\co}{covariate\xspace}

\newcommand{\sti}{stopout\xspace}

\newcommand{\selfself}{self-proposed, self-extracted\xspace}
\newcommand{\crowdself}{crowd-proposed, self-extracted\xspace}
\newcommand{\crowdcrowd}{crowd-proposed, crowd-extracted\xspace}

\newcommand{\neither}{passive collaborator\xspace}
\newcommand{\wiki}{wiki contributor\xspace}
\newcommand{\forum}{forum contributor\xspace}
\newcommand{\both}{fully collaborative\xspace}
\newcommand{\Si}{Simple\xspace}
\newcommand{\Co}{Complex\xspace}
\newcommand{\edx}{edX\xspace}
\newcommand{\De}{Derived\xspace}

\newcommand{\ile}{\textit{lead}\xspace}
\newcommand{\la}{\textit{lag}\xspace}

\newcommand{\tten}{average pre deadline submission time\xspace}
\newcommand{\tnine}{correct submissions percent\xspace}

\newcommand{\tseven}{lab grade over time\xspace}
\newcommand{\tsix}{lab grade\xspace}
\newcommand{\tfive}{pset grade over time\xspace}
\newcommand{\tfour}{pset grade\xspace}
\newcommand{\tthree}{average number of submissions in percent\xspace}

\newcommand{\feleven}{submissions per correct problem\xspace}
\newcommand{\ffive}{average length of forum post\xspace}

\newcommand{\MITX}{MITx\xspace}

 \usepackage[accepted]{icml2013}

\begin{document} 

\twocolumn[
\icmltitle{Towards Feature Engineering at Scale \\for Data from Massive Open Online Courses}


\icmlauthor{Kalyan Veeramachaneni}{kalyan@csail.mit.edu}
\icmladdress{Massachusetts Institute of Technology,
         Cambridge, MA 02139 USA}
\icmlauthor{Una-May O'Reilly}{unamay@csail.mit.edu}
\icmladdress{Massachusetts Institute of Technology,
           Cambridge, MA 02139 USA}
\icmlauthor{Colin Taylor}{colint@csail.mit.edu}
\icmladdress{Massachusetts Institute of Technology,
           Cambridge, MA 02139 USA}

\icmlkeywords{boring formatting information, machine learning, ICML}

\vskip 0.3in
]

\begin{abstract} 
We examine the process of engineering features for developing models that improve our understanding of learners' online behavior in MOOCs.  Because feature engineering relies so heavily on human insight, we argue that extra effort should be made to engage the crowd for feature proposals and even their operationalization.  We show two approaches where we have started to engage the crowd. We also show how features can be evaluated for their relevance in predictive accuracy. When we examined crowd-sourced features in the context of predicting stopout, not only were they nuanced, but they also considered more than one interaction mode between the learner and platform and how the learner was *relatively* performing.  We were able to identify different influential features for stop out prediction that depended on whether a learner was in 1 of 4 cohorts defined by their level of engagement with the course discussion forum or wiki.
This report is part of a compendium which considers different aspects of MOOC data science and stop out prediction. 
\end{abstract} 
\section{Introduction}

\begin{figure*}
\centering
    \includegraphics[width=1\textwidth]{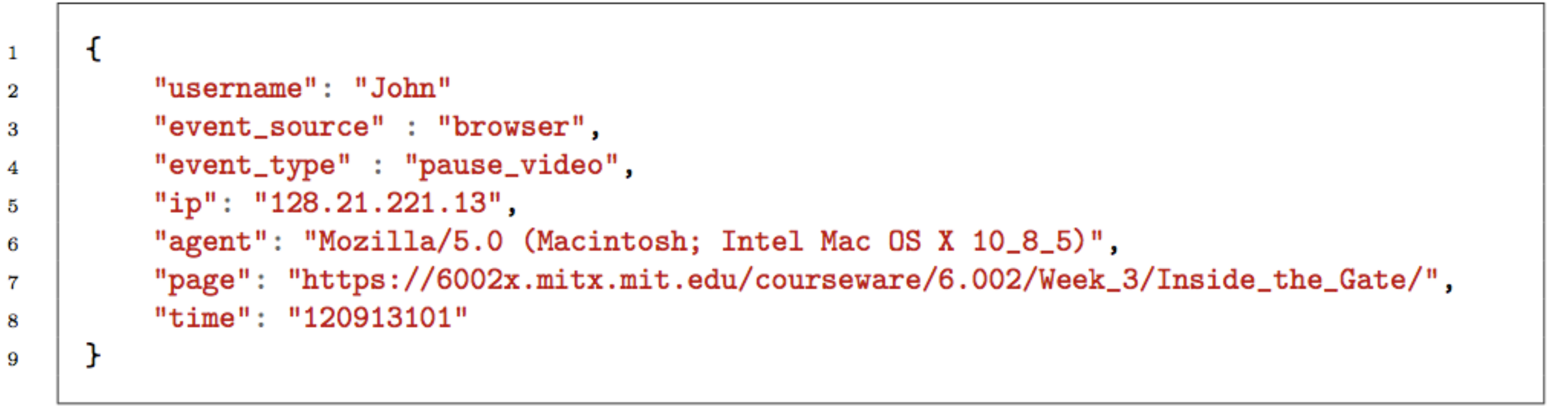}
\caption{An example log line stored in the backend as a learner pauses a video. The log line is stored in JavaScript Object Notation (JSON).} 
\label{json-example}
\end{figure*}

Massive open online courses (MOOCs) have provided a new way to reach enormous numbers of learners via an online platform. Many courses proceed very similarly to a campus course. However, they allow students from anywhere in the world to register for a course and earn a certificate upon successful completion. The specific layout of each MOOC varies, but currently most follow a similar format. Content is sectioned into modules, usually using weeks as intervals. Most MOOCs include online lectures (video segments), lecture questions, homework questions, labs, a forum, a Wiki, and exams (midterm and final). Students \footnote{In this paper we use learner and student to mean the same} advance through the modules sequentially, access online resources, submit assignments and participate in peer-to-peer interactions (forums).  While similar to campus based courses there are significant differences in the way material is offered and the way learners interact with these courses. \footnote{A TEDx lecture by Anant Agarwal explains these and other key ideas that make MOOCs powerful}: 
\vspace{-3mm}
\begin{description}
\item \textbf{Interactive learning}: MOOCs allow the insertion of interactive exercises in between lecture videos enabling student to apply the concepts they have learnt. They allow instructors to use new technology to engage learners in course material including simulations, peer grading, discussion forums, and  games. They also allow instructors to integrate experiences from outside classroom into the curriculum. 
\vspace{-3mm}
\item \textbf{Self paced and anytime learning}: They allow students to start their lectures anytime and engage with the course anytime as and when their schedule permits. Additionally, students can replay lecture videos, pause, play again and learn at a pace that is most beneficial for them. 
\vspace{-3mm}
\item \textbf{Instantaneous feedback on assessments}: In MOOCs students can be allowed multiple attempts for a problem and can get instantaneous feedback on every attempt. This feedback can range from whether the answer was right or wrong to a more sophisticated diagnosis. 
\end{description}
\vspace{-4mm}

\begin{figure}
\texttt{
 \begin{tabular}{lll}
  timestamp & url & event \\ \hline
 $2013-11-10 08:46:21$& 191  & play\_video   \\
  $2013-11-10 08:46:49$ & 191 & pause\_video \\
   $2013-11-10 08:47:24$& 191 &  play\_video  \\
   $2013-11-10 08:51:25$ & 191 & pause\_video  \\
  $2013-11-10 08:51:48 $& 191 &play\_video \\
   $2013-11-10 08:53:08 $&198 & seq\_goto   \\
   $2013-11-10 08:55:05$& 284 & pause\_video\\
   $2013-11-10 08:56:05 $& 284 &play\_video\\
   $2013-11-10 09:40:50 $&284 &pause\_video \\
   $2013-11-10 09:41:13 $&284& play\_video \\
    $2013-11-10 09:41:57$&284&play\_video \\
    $2013-11-10 09:53:37$&284& pause\_video \\
    $2013-11-10 10:15:53$& 284& problem\_check \\
    $2013-11-10 10:20:27$ & 121 &problem\_check  \\
    $2013-11-10 10:22:27$& 123 & problem\_check  \\
    $2013-11-10 10:25:50$ &123  & problem\_graded \\
 \end{tabular}
}
\caption{A snapshot of one learner's timeline spanning approximately 2 hours of activity recorded as click stream events. During this period the student \textit{plays} and \textit{pauses} a video and attempts the problems available on the \textit{urls} 121 and 123. \textit{urls} are encoded with numbers and we store all the meta information about the content of the \textit{url} in a different table.} 
\label{studentevents}
\end{figure}

While students advance in the course, every mouse click they make on the course website is recorded, their submissions are collected and their interactions on forums are recorded. An example of a clickstream event and how it is recorded is presented in Figure~\ref{json-example}. The recorded data presents an opportunity for researchers to analyze the data post-hoc, and answer questions ranging from simple ones such as \textit{what was useful?} and \textit{what was not?} to more complex research questions such as \textit{What was the reason behind a student leaving the course?}, \textit{What were the most common misconceptions in the material?}, \textit{How do students solve a problem?}. 


As data scientists, to answer these questions one first attempts to quantitatively characterize learners online behavior from \textit{web logs} and \textit{click stream} data. The raw data, recorded as shown in the Figure~\ref{json-example},  after processing, curation, and storing in a database \footnote{These three steps are extremely complex and challenging but are not in the scope of this paper}, enables extraction of \textit{per-learner} sequences of click stream events during a specified time period, shown in Figure~\ref{studentevents}. These \textit{per learner} sequence of the events only provide primitive signals that form bases for inferences such as learner's knowledge acquisition, attitude, attention. However, they have potential to help us gauge learners intent, interest and motivation in the absence of verbalized or visual feedback from the learner. To make such inferences we must form \textit{variables} capturing learner's behavior; an endeavor we call \textit{feature engineering}. Among many different types of \textit{variables} and data representations, two kinds of variables are of interest: 

\noindent \textbf{Variables that capture per learner behavior with respect to a \textit{resource}}: For example, in the above sequence of clickstream events two such variables are: \textit{total time spent of the video} and the \textit{number of pauses while watching the video}. When these two variables are evaluated for all the learners and analyzed they can uncover patterns; if too many learners \textit{pause} too many times, the video could be fast and/or confusing.  

\noindent \textbf{Per-learner longitudinal variables}: A longitudinal study involves repeated observation of the same variables over time. A variable is usually an aggregate or a statistic of some observations defined for that time period. In the context of the MOOC, we can define the time interval to be a \textit{week}, a \textit{day} or a \textit{time corresponding to the module/unit} or divide the course into two periods - before and after midterm. An example is \textit{time each student spent on the course website during the week}. A more complex variable is \textit{on an average, the time before the deadline learner starts to work on an assignment}.

\subsection{What is the challenge in feature engineering?}
Engineering features from this type of data: time series of click stream events that record human interaction with an online learning platform presents a very unique set of challenges. As machine learning researchers (and data scientists) our first inclination was to seek automated ways to extract features. Perhaps the methods developed for problems in image understanding, and both text and spoken language processing, such as \textit{deep learning} that enable further automation of what has  already been semi-automated would transfer? With this type of data, however, we quickly realized that feature engineering needs to be primarily driven by humans because of the multiple roles they assume in this endeavor.  Below we explicate through examples some of the roles humans play in engineering these features. They: 
\vspace{-3mm}
\begin{description}
\item \textbf{Generate ideas based on their intuition}: Coming up with variables  requires generation of ideas based on \textit{intuition},  and understanding of what could be relevant for a study. As humans who have been learners in some shape or form, we self reflect to invent variables.  For example, when considering prediction of stopout/dropout, we might each quite naturally suggest  ``\textit{If the student starts to homework problems very close to the deadline, he might be very likely to fall behind and eventually drop out}". Subsequently, we might propose how to operationalize such a variable into a quantitative value by measuring, `` Time difference between the dead line and the students first attempt for the homework problem".  While many other aspects of feature engineering can be automated, this one cannot. 
\vspace{-3mm}
\item \textbf{Bring their knowledge about the context as instructors}: For MOOCs, designing variables requires understanding of \textit{context} and \textit{content} of the course for which the variables are sought. In other words instructors or experts in the course are able to propose what variables might be important to capture. For example, an instructor might be aware of an important concept whose understanding is critical for continued success in the course and may hypothesize that a variable that captures whether the learner understood the concept or not could help predict stopout/dropout. 
\item \textbf{Use their highly specialized knowledge of learning science}: Additionally researchers from learning sciences are able to propose variables grounded in theory that they can link together \textit{via} a multivariate distribution to explain latent constructs such as \textit{motivation}, \textit{intention}, and \textit{self-efficacy}. 
\item \textbf{Operationalize the ideas}: Due to the type of data and a nature of variables, operationalizing these ideas into variables require a number of steps, addition of details depending upon the context, assumptions and heuristics. Take for example the variable proposed above. First it requires us to assemble the deadlines for all the problems in different weeks. Then we have to define as to what constitutes as ``start"  time for student working on the assignment. Since there is no mechanism where students notify when they started to work on the assignment, we have two options: the first time they looked at the problem, or the time of the first attempt for the problem or the time they attempted but saved the answer instead of checking for correctness. One can argue that the first time student looks at the assignment might not be the start time as it might correspond to the learners simple browsing behavior, so one resorts to the first attempt made by the learner towards the problem. 
\end{description} 
\vspace{-3mm}

Given that humans are NOT replaceable in this endeavor, we shifted our focus towards addressing a different goal: \textit{how to increase the number of people who can participate in this endeavor?}. Towards that, in this paper, we initiate a new, broad and fundamental approach towards human driven feature engineering. In developing our new approach we focused on engineering features that could be predictors for \textit{who is likely to stopout?}   
 
We started with a very natural approach which was to think up feature ideas ourselves and transfer them into quantities in the data. We realized that this tact is vulnerable to missing some features because there are other interpretations of what was happening that could be different from ours. This led us to construct activities which involve collecting ideas for features from others, i.e. the "crowd". This allowed us to expand the diversity of our set and eliminate our blind spots. Subsequently we evaluated the value of features in predicting \sti from a machine learning perspective and discerned an important characterization of features. Our study on feature engineering for stopout prediction problem helped us lay the foundation for building next generation scalable feature engineering platforms that are not only able to radically increase the number of people who can participate in this endeavor, but also enable a much smoother and efficient participation. 

\subsection{Our contributions}
Our contributions through this paper are: 
\begin{itemize}
\vspace{-2mm}
\item We develop the first set of longitudinal \textit{per-learner} variables that are able to successfully predict \sti. 
\vspace{-2mm}
\item For the first time, in this domain, we present the results of soliciting ideas for variables from the ``crowd", an endeavor we conclude is going to be necessary in this domain.  
\vspace{-2mm}
\item We provide an in-depth account of the feature engineering process we have used to develop features that helped us predict \sti.\footnote{Stopout is what we refer to for dropout}
\vspace{-2mm}
\item We present a systematic way to evaluate the contribution of each feature towards the prediction problem. We apply this methodology to the features we assembled and demonstrate the importance of the contribution of the ``crowd". 
\vspace{-2mm}
\item We present the insights feature engineering can yield about feature types, and how to make features themselves as carefully nuanced and developed as the predictive models of behavior.
\vspace{-2mm}
\item We use our account to reflect on the feature engineering that is going to be necessary if the community wants to fulfill the goal of understanding online learning behavior from MOOC data. We present the steps necessary to scale this process and increase the pool of people contributing to finding insights from the data.
\end{itemize}

\begin{figure*}[ht!]
\centering
    \includegraphics[width=0.7\textwidth]{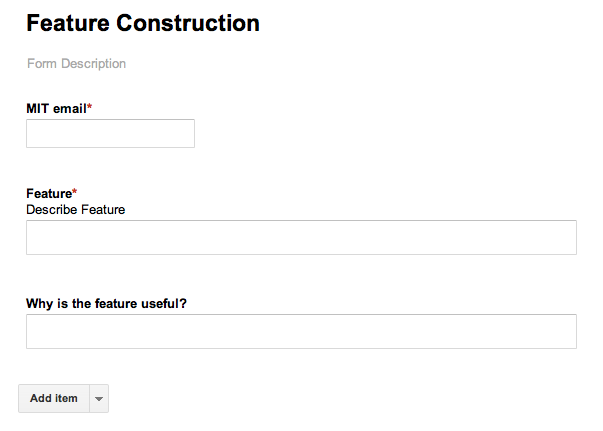}
      \caption{Google form presented to the members of the class. The participants were asked to describe the \textit{feature} they propose, and why it would be useful in predicting \sti. A total of 30 \textit{features} were proposed, by 12 members in the class. 8 members proposed more than 1 \textit{feature}.}\label{fig:googleform}
\end{figure*}
 
\setlength{\arrayrulewidth}{0.5pt}
\setlength{\tabcolsep}{10pt}
\renewcommand{\arraystretch}{1.5}

\newlength{\thickline}
\setlength{\thickline}{1pt}
\makeatletter
\def\hlinex{%
  \noalign{\ifnum0=`}\fi\hrule \@height \thickline \futurelet
   \reserved@a\@xhline}
\makeatother

\newlength{\threecoltabwid}
\setlength{\threecoltabwid}{\textwidth - \tabcolsep * 2 * 3}

\begin{table*}[htp]
 \centering
 \medskip
\begin{tabular*}{\textwidth}{ >{\centering\arraybackslash}p{0.05\threecoltabwid} >{\raggedright\arraybackslash}p{0.35\threecoltabwid} >{\raggedright\arraybackslash}p{0.6\threecoltabwid}}
 \hlinex
& \textbf{Describe feature}      & \textbf{Why is this feature useful?}          \\ \hlinex
  & \textbf{pset grade over time}: Difference between grade on the current pset and average grade over previous psets. Significant decreases may be likely to be correlated with dropouts.& Anecdotally it appears that users who perform poorly on the current week (especially after not performing poorly in the preceding weeks) will subsequently give up. They may also, with low probability, post on the forum explaining their issue with the task at hand.    \\ \hline
 
&\textbf{average pre deadline submission time}: average time between problem submission time and problem due date. & people who get things done early are probably not under time pressure that would make them drop out. 
 \\ \hline
&\textbf{proportion of time spent during weekends)}: Fraction of observed resource time spent on each day of the week (7 variables for Mon-Sun that add up to 1). Just for previous week, and averaged over all weeks so far.  & Heavy weekend users might be more likely to drop out, because they don't have spare weekday time to devote to the course. 
           \\ \hlinex
\end{tabular*}
 \caption{Three examples of features proposed by the students and instructors in the MIT class.}\label{cf}

\end{table*}

\vspace{-3mm}

We proceed in the following manner. In Section~\ref{sect:course} we start by describing the specific MOOC data we are working with. In Section~\ref{sect:ideation} we present 3 approaches to initially proposing features. In Section~\ref{sect:cf} we elaborate upon one of these: crowd sourcing. Next in Section~\ref{sect:features} we move to describing the operationalization of the original feature ideas and proposals and list the features we eventually engineered. We discuss, in Section~\ref{sect:challenges} the types of features and the challenges of operationalization. In Section~\ref{sect:ml}, we describe how we evaluated the importance of these features with predictive modeling based upon machine learning. In Section~\ref{sect:related} we present other efforts of feature engineering in this domain and compare with ours. We conclude in Section~\ref{sect:conclusions} with a summary of our findings and our next steps. 
\section{Data collected}\label{sect:course}
We engineered features related to stopout  from data collected from a MOOC offered on the \MITX platform\footnote{\MITX became what is now known, circa 2013, as \edx}. The course is 6.002x: Circuits and Electronics taught in Fall of 2012. 6.002x had 154,763 registrants. Of those, 69,221 students looked at the first problem set, and 26,349 earned at least one point. 9,318 students passed the midterm and 5,800 students got a passing score on the final exam. Finally, after completing 15 weeks of study, 7,157 registrants earned the first certificate awarded by MITx, showing they had successfully completed 6.002x.

edX provided the following raw data:
\vspace{-2mm}
\begin{itemize}
\item A dump of click-stream data from student-browser and edX-server tracking logs in JSON format. For instance, every page visited by every student was stored as a server-side JSON (JavaScript Object Notation) event.
\vspace{-2mm}
\item Forum posts, edits, comments and replies stored in a MongoDB collection. Note that passive forum activity, such as how many views a thread received was not stored here and had to be inferred from the click-stream data.
\vspace{-2mm}
\item Wiki revisions stored in a MongoDB collection. Again, passive views of the Wiki must be inferred from the click-stream data.
\vspace{-2mm}
\item A dump of the MySQL production database containing student state information. For example, the database contained his/her final answer to a problem, along with its correctness. Note that the history of his submissions must be inferred from the click-stream data.
\vspace{-2mm}
\item An XML file describing the course calendar which included information like the release of content and the assignment deadlines.
\end{itemize}
\vspace{-3mm}
This data included 17.8 million submission events, 132.3 million curated navigational events \footnote{We received more navigational events, but only 132.3 million were well formed enough to be reliably considered for this paper. } and 90,000 forum posts. 

To analyze this data at scale, as well as write reusable feature engineering scripts, we first organized the data into a schema designed to capture pertinent information. The database schema, MOOCdb, is designed to capture MOOC data across platforms thereby promoting collaboration among MOOC researchers. MOOCdb utilizes a series of  scripts to pipe the raw data into a standardized schema.  It identifies 4 basic student-platform interaction modalities: observing, submitting, collaborating and giving feedback.  In observing mode students (somewhat passively) browse the online platform, watch videos, read material such as e-books or examine forum posts. In submitting mode, students submit information to the platform such as quiz responses, homework solutions, or other assessments. In collaborating mode students post to other students or instructors on forums, add material to a wiki or chat on google hangout or other social venues. In feedback mode students respond to surveys. MOOCdb encompasses and organizes the detailed data collected during these modalities. Its aim is to be platform agnostic by means of providing a common terminology between platforms. More about MOOCdb can be found in the MOOCdb Tech report, but the details about the schema itself are outside the scope of this report \cite{tr}.  
\section{Our approaches for feature ideation}\label{sect:ideation}
With the database, we then proceeded to form ideas for the \fes that we can repeatedly calculate on a \textit{per-student} basis. We proceeded in three different ways:
\vspace{-2mm}
\begin{itemize}
\vspace{-2mm}
\item \textbf{Approach 1}: We brainstormed feature ideas ourselves. Next, we operationalized our own ideas by writing feature extraction scripts. We call these features \selfself.
\vspace{-2mm}
\item \textbf{Approach 2}: We asked others for ideas of what might be predictive of \sti. The people we asked included students, teachers and other researchers. We refer to this group collectively as `the crowd.' We identified ideas that we had not implemented yet, and constructed feature extraction scripts ourselves. We call these \crowdself. In the next section we provide more details for this approach. 
\vspace{-2mm}
\item \textbf{Approach 3}:  Finally, we asked `the crowd' to brainstorm predictive features, and to send us feature extraction scripts that we could run on MOOCdb. We provided people with a mock dataset with an identical data schema. Thus, instead of providing actual student data, we empowered the crowd to join in our data science efforts. We call the resulting features \crowdcrowd.
\vspace{-2mm}
\end{itemize}
\vspace{-3mm}
Below we present the crowd sourcing experiment we performed in approach 2. 

\section{Approach 2: Crowd sourcing}\label{sect:cf}
To generate ideas for \fes, we sought opinions from a class at MIT. We presented the data model (what was being collected), explained what we meant by a \fe and asked members of the class (professors and students) to posit features for each student that could predict a student's \sti. We collected the input \textit{via} a google form presented in Figure~\ref{fig:googleform}. In this form we asked the users to describe the \fe and describe why they think the \fe will be useful in predicting \sti?. We did not present our features to the class.  

\setlength{\arrayrulewidth}{0.5pt}
\setlength{\tabcolsep}{10pt}
\renewcommand{\arraystretch}{1.5}

\begin{table*}[htp]
\centering
\begin{threeparttable}
 \caption{List of \selfself \cov}\label{table:self_proposed_self_extracted}
\medskip
  \begin{tabular*}{\textwidth}{ >{\centering\arraybackslash}p{0.05\threecoltabwid} >{\raggedright\arraybackslash}p{0.25\threecoltabwid} >{\raggedright\arraybackslash}p{0.7\threecoltabwid}}
\hlinex
& \textbf{Name}  & \textbf{Definition}  \\ \hlinex
\x{1}  & stopout & Whether the student has stopped out or not  \\ \hline
*\x{2} & total duration& Total time spent on all resources  \\ \hline
\x{3}  & number forum posts  & Number of forum posts\\ \hline
\x{4}  & number wiki edits& Number of wiki edits\\ \hline
*\x{5} & average length forum post& Average length of forum posts\\ \hline 
*\x{6}  & number distinct problems submitted & Number of distinct problems attempted \\ \hline 
*\x{7} & number submissions  & Number of submissions \tnote{1}\\ \hline
\x{8}  & number distinct problems correct & Number of distinct correct problems \\ \hline 
\x{9} & average number submissions & Average number of submissions per problem (\x{7} / \x{6})\\ \hline 
\x{10} & observed event duration per correct problem  & Ratio of total time spent to number of distinct correct problems (\x{2} / \x{8}). This is the inverse of the percent of problems correct \\ \hline 
\x{11} & submissions per correct problem  & Ratio of number of problems attempted to number of distinct correct problems (\x{6} / \x{8}) \\ \hline
\x{12} & average time to solve problem & Average time between first and last problem submissions for each problem (average(max(submission.timestamp) - min(submission.timestamp) for each problem in a week) )\\ \hline
*\x{13} & observed event variance& Variance of a student's observed event timestamps    \\ \hline
\x{14} & number collaborations& Total number of collaborations (\x{3} + \x{4})   \\ \hline
\x{15} & max observed event duration & Duration of longest observed event  \\ \hline
*\x{16} & total lecture duration& Total time spent on lecture resources \\ \hline
*\x{17} & total book duration & Total time spent on book resources  \\ \hline
*\x{18} & total wiki duration & Total time spent on wiki resources  \\ \hlinex
  \end{tabular*}
 \medskip
\begin{tablenotes}
\footnotesize
\item[1] In our terminology, a submission corresponds to a problem attempt. In 6.002x, students could submit multiple times to a single problem. We therefore differentiate between problems and submissions.
\end{tablenotes}
\end{threeparttable}
\end{table*}
\noindent \textbf{Outcomes}: Out of the 30 features that the class proposed, 7 were in common with ours. Out of the remaining 23 features, we extracted 10 features. These features are listed in Table~\ref{table:crowd_proposed_self_extracted} and are listed with numbers starting from 200.  

The features proposed by the students and instructors in this class were \textit{intuitive}, based on \textit{experience} and self identification as once/or currently being a student. Participants also gave detailed reason as to why the feature is useful. We present three examples in Table~\ref{cf}.  

\section{Operationalizing features ideas/proposals}\label{sect:features}
After curating the data and carefully gathering the proposals for features, we started operationalizing these hypothesized to be predictive of \sti. We split the course into 15 time slices/weeks. Thus, for each defined feature, we assigned each student a feature-value each week. For example, each student has a value for the feature, \textit{number of forum posts}, for each of the 15 weeks. For each week, we also assign a value for \sti. The value is 0 if the student has already stopped out by not submitting any more assignments, or it is 1 if the student will submit assignments in the future.
\subsection{Self-proposed, self-extracted features}
Table \ref{table:self_proposed_self_extracted} summarizes the features we completely developed ourselves. Each feature is calculated on a per student, per week basis. For features with *, additional details are necessary because how they are operationalized is ambiguous and we made several decisions while operationalizing them. 
\begin{itemize}
\item \x{2}, \x{16}, \x{17}, \x{18}: These features are based on observed event duration. The edX server logs did not explicitly provide this, so we need to infer the duration based on the timestamps of the start of observed events. We assume that a student observed an event until he observed a different event (a new timestamp). This is a similar approach used by industry web-profile metrics. For example, if Student A had three observed events with timestamps, T1, T2 and T3, the duration of the first event would be T2 - T1, the duration of the second is T3 - T2.  Sometimes, the spacing between observed events is very large, presumably because the user stopped interacting with the website.  This is handled by setting the last observed event's duration to a MAX\_DURATION. Hence if $T3 - T2 > 60$, the duration is set to MAX\_DURATION. In our case, we set MAX\_DURATION to be 60 minutes, because our data included durations of up to $\sim$ 60 minutes , Additionally, the duration of the third event is MAX\_DURATION, since there is no T4. 
\item \x5: A forum post's length is the number of characters in the forum post (i.e. the length of the string). We used MySQL's length function.
\item \x6, \x7: With problem submissions, week number is ambiguous. Students may submit a problem at any time (assuming the problem is released), regardless of when the problem is due. In other words, even if a problem corresponds to week number 3, a student could submit that problem in week 5. For these features, we counted a submission in week w1 if the submission's timestamp is in w1, regardless of whether or not the problem is part of w1's assigned content.  We chose to do this because the feature is meant to capture a student's weekly activity.
\item \x{13}: For this feature, we tried to measure the consistency of a student's observed event patterns relative to the time of day (i.e., a student who always works on the course at 7:00 a.m. would have small variance for that week). To capture event variance, for each day, we counted the number of seconds after midnight of the observed event timestamp. We created a distribution of all of the number of seconds for each student each week. Then, we calculated the variance of the distribution (each student, week pair has it's own distribution). This variance becomes the feature. Note: student's participate from around the world, but the timestamp is in UTC time. However, because variance is valued over absolute value, the actual time is irrelevant. 
\end{itemize}
\setlength{\arrayrulewidth}{0.5pt}
\setlength{\tabcolsep}{10pt}
\renewcommand{\arraystretch}{1.5}

\begin{table*}[htp]
 \centering
 \caption{List of \crowdself \cov}\label{table:crowd_proposed_self_extracted}
 \medskip
\begin{tabular*}{\textwidth}{ >{\centering\arraybackslash}p{0.05\threecoltabwid} >{\raggedright\arraybackslash}p{0.35\threecoltabwid} >{\raggedright\arraybackslash}p{0.6\threecoltabwid}}
 \hlinex
   & \textbf{Name}      & \textbf{Definition}          \\ \hlinex
 $x_{201}$ & number forum responses & Number of forum responses      \\ \hline
 *$x_{202}$ & average number of submissions percentile & A student's average number of submissions (feature 9) as compared with other students that week as a percentile  \\ \hline
 *$x_{203}$ & average number of submissions percent & A student's average number of submissions (feature 9) as a percent of the maximum average number of submissions that week             \\ \hline
 *$x_{204}$ & pset grade    & Number of the week's homework problems answered correctly / number of that week's homework problems                \\ \hline
 $x_{205}$ & pset grade over time   & Difference in grade between current pset grade and average of student's past pset grade                   \\ \hline
 *$x_{206}$ & lab grade    & Number of the week's lab problems answered correctly / number of that week's lab  problems                \\ \hline
 $x_{207}$ & lab grade over time  & Difference in grade between current lab grade and average of student's past lab grade                   \\ \hline
 $x_{208}$ & number submissions correct  & Number of correct submisions      \\ \hline
 $x_{209}$ & correct submissions percent  & Percentage of the total submissions that were correct ($x_{208}$ / $x_{7}$)     \\ \hline
 *$x_{210}$ & average predeadline submission time & Average time between a problem submission and problem due date over each submission that week               \\ \hlinex
\end{tabular*}

\end{table*}
\subsection{Crowd-proposed, self-extracted features} \label{section:crowdself}
Table \ref{table:crowd_proposed_self_extracted} summarizes the features the crowd hypothesized, but we extracted. Each feature is calculated on a per student, per week basis. A * indicates that a disambiguating explanation follows underneath.
\begin{itemize}
\item \x{202}, \x{203}: For each week, we create a distribution of all of the values for every student of feature \x9. Then, we compare a student's \x9 value to the distribution for that week. \x{202} is the percentile over that distribution, and \x{203} is the percent as compared to the max of the distribution.
\item \x{204}, \x{206}: As mentioned earlier, with regard to submissions, there is an ambiguity: whether a submission correspond to the week in which it was submitted, or the week in which the problem's module was. These features are meant to capture the grade on the module. Therefore, they are computed based on the week's homework assignment and lab assignment, rather than on the submission timestamp. The number of problems the student answered correctly out of the total number of homework or lab problems corresponding to that week constitute features \x{204} and \x{206}.
\item \x{210}: For each submission during the week, the time difference between the submission timestamp and the due date of the problem is calculated. \x{210} is the average of all of these differences. 
\end{itemize}

\subsection{Crowd-proposed, crowd extracted features}
In an attempt to crowdsource feature extraction, we asked SQL-fluent MIT students and researchers to both hypothesize new features and submit scripts which would extract them. We are still in the process of collecting feature scripts from this effort at the time of writing. 

\section{Types of features and challenges}\label{sect:challenges}
 Our 28 features are more sophisticated because of the variety of sources used in their proposition, such as the leveraging crowd-sourced brainstorming  to capture creative behavioral features. Many involve complexities beyond a simple count per week. Such complexities include:
\begin{description}
\item \textbf{Use of higher level statistics} We use, for example, the variance of the times of day that a student accesses course material each week (\x{13}) and the percentile of a student's average number of submissions (\x{202}). (\x{202}) also is a relative standing of the student amongst his peers. 
\item \textbf{Curation requiring human cross-referencing}: Some features required manual curation in order to arrive at a descriptive metric. For example, \x{204}, a student's \textit{pset} grade, necessitated manual curation of problem and assignment deadlines from the course content. 
\item \textbf{Referencing multiple data sources and MOOCdb modes}: Some features required linking information from multiple sources (such as \x{204}, the pset grade). This included getting deadlines for the problems from the XML file, all submissions from the server logs, and the problem's correctness from the production MySQL dump.
\item \textbf{Computationally expensive processing}: Because the features are defined on a \textit{per-student} \textit{per-week} basis, we must extract events on a \textit{per-student} basis for every week and then extract information for each student for that week. With millions of events in the database and hundreds of thousands of students this processing is computationally expensive. 
\item \textbf{Integration of human intuition and context}: Some features express subtle human intuition about motivational context. For example, \x{10} represents the amount of time a student spends on the course (\x{2}) per \textit{correct} problem (\x{8}). This feature captures the less tangible gratification a student experiences based on time spent.
\end{description}
These observations have prompted us to discern 3 feature categories; we explain them by the way of examples: 
\begin{description}
\item \textbf{\Si}: These features require a simple count or creation of an aggregate for every student on a per week basis. The count or aggregate is usually over an already existing field or a count of a certain type of events. Examples include: \textit{total time spent on the course}, \textit{number of problem attempts made during this week}, and \textit{amount of time spent on the videos (or a certain video)}. 
\item \textbf{\Co}: These features require extraction of relational linking of data from two or more modes of student interaction. This implies more complex processing (and pre processing to link the data). They may require curation and some additional manual processing. Examples of these features include: \textit{number of times the student goes to forums while attempting problems}, or \textit{on an average how close to the deadline does the student start attempting problems}, and \textit{observed event duration per correct problem}. 
\item \textbf{\De}: These features combine one or more simple or complex features to form a new feature. Usually a mathematical function, like ratio, trend, or percentile is used in the composition. An instructor or student familiar with the course brings some domain expertise to propose such a feature. Essentially human intuition plays a key role. Examples of these type of features include: \textit{ratio of number of distinct problems correct to the total time spent on the course during that week}, \textit{the difference in pset grade of the current week and the average pset grade over past weeks} and \textit{a learner's number of submission events in percentile (compared against his peers)}. 
\end{description}
 
\section{Evaluating our features}\label{sect:ml}
To evaluate our \fes in terms of how well they collectively explain stopout, we use them in a supervised learning scenario we next describe. 

\subsection{Supervised learning problem: \sti prediction}
\begin{figure*}[!ht]
  \caption{Diagram of the students' weeks data used in a lead 5, lag 3 prediction problem}\label{fig:lead_lag}
  \centering
    \includegraphics[width=0.8\textwidth]{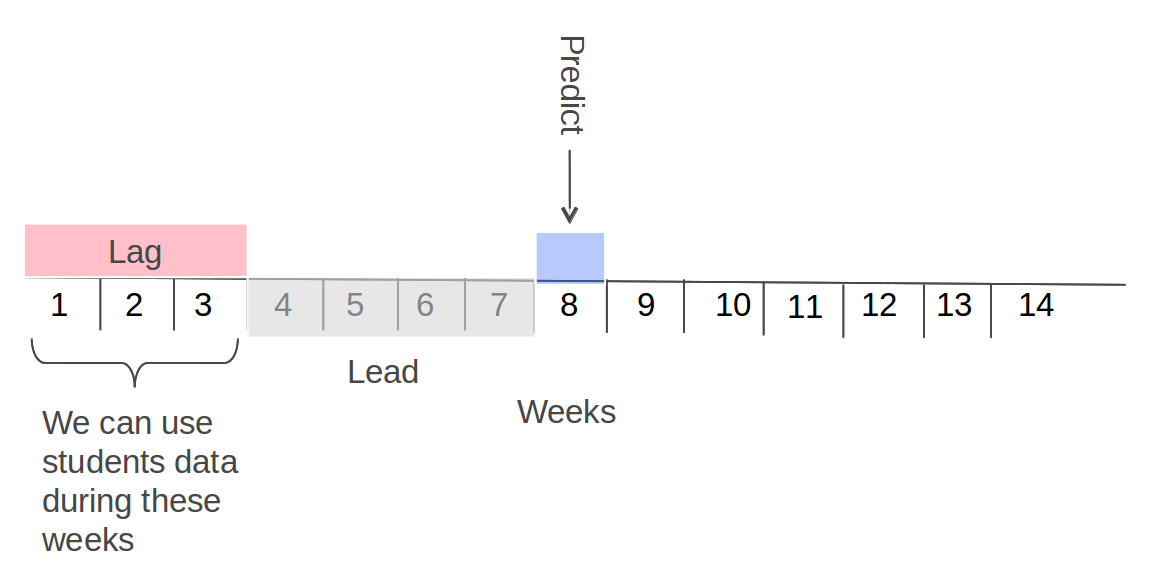}
\end{figure*}

Our goal is to use our features to predict \sti. We consider a student to have stopped out if s/he stops attempting the problems in the course. In this prediction problem, based on students behavior up until a time point (using the historical data up until that time point, a.k.a \textit{lag}), we predict whether or not a student will \sti by a certain time in future separated by a time interval from the current time point (called \ile). Thus \ile represents how many weeks in advance we attempt to predict \sti. For example, if we use a \ile of 5 and a \la of 3, we would take the first 3 weeks of data to predict 5 weeks ahead, that is predict for the 8th week. Thus, each training data sample consists of repeated measurements of student's feature values for weeks 1, 2 and 3 as \cov. The binary \sti value for week 8 becomes the label. Figure \ref{fig:lead_lag} shows a diagram of this scenario.

We are careful not to use students' stopped out week's features as input to our models. In other words, if a student has stopped out in week 1, 2 or 3, we do not use this student as a data point. Including stopped out student data makes the classification problem too easy as the model will learn that a stopped out student never returns (by our \sti definition).

Since there were 14 weeks in the course, we have a total of 91 unique prediction problems by varying \ile and \la. We consider the above example as one prediction problem.

\noindent \textbf{Randomized Logistic Regression}
We use randomized logistic regression to assess the importance of features. Our model uses 27 features to model \sti. In order to best fit a training set, a logistic regression model optimizes weights for each feature.  To assess the importance of the features randomized logistic regression repeatedly models a perturbed data set (subsample) with regularization, and  works as follows: 

\begin{description}
\item {Step 1:} Sample without replacement  75\% of the training data each time (the variables are normalized ahead of training). 

\item {Step 2:} Train a logistic regression model on the sub-sampled data, with randomized regularization coefficient for each variable. The randomized coefficient $\beta_j$ is sampled from uniform distribution $[\lambda, \frac{\lambda}{\alpha}]$, where $\alpha \in (0,1]$ and $\lambda$ is the regularization coefficient usually used in standard regularized regression approaches. This randomization places different selection pressure for different variables. 

\item {Step 3:} For every \co evaluate $b_s^{j}=\mu(w_j,th)$ where $\mu$ is a unit step function and $w_j$ is the \coeff for \co $i$ and $th$ is the threshold we set to deem the feature important. This is set at 0.25.  Thus this results in a binary vector, that represents the selection of the covariate. This binary vector is ($lag \times |features|$) long where $1$ at a location $j$ implies feature $i= j \mod 27$ was present in this model. 

\item {Step 4:} Repeat Steps 1, 2 and 3 a total of 200 times. 

\item {Step 5:} Estimate the importance of the \co $j$ by calculating the selection probabilities $\frac{\sum_s b_s^{j}} {200}$. 

\end{description}

\subsection{Experimental setup and results}
We divided our learners into four cohorts. These are \neither, \wiki, \forum and \both. Learners who did not post in forums (but may have visited forums) and did not edit wiki pages were categorized as \neither. Learners who participated in forums but not wiki were categorized as \forum, and who edited wiki but did not participate in forums were categorized as \wiki and learners who participated in both were categorized as \both. We ran randomized logistic regression for every \ile, \la and cohort combination. Thus we have run $91 \times 4$ randomized logistic regression experiments (an experiment is described above). In each experiment 200 logistic regression models are formed, thus adding up to a total of approximately 72,000 logistic regression models.  For each experiment, randomized logistic regression resulted in a vector of \cov selection probabilities. Each of these probabilities ranged from 0 to 1. \footnote{We used the scikit-learn Randomized Logistic Regression implementation.}

Randomized logistic regression analysis gave us fascinating covariate selection probability vectors for all 91 experiments and all cohorts. For each experiment the randomized logistic regression gives us these selection probability vectors for all the \cov which are learner features for different weeks.  In order to gain a more quantitative grasp of which features matter for different prediction problem, we aggregate these probabilities. 

\begin{paragraph}
{Week invariant feature importance} To calculate the importance of a feature for each cohort we follow the two steps: 
\vspace{-2mm}
\begin{description}
\item (1) We first evaluate its importance in an experiment by summing its selection probability across different weeks. We then divide this sum with the \la for that experiment. This is a heuristic which gives the feature's importance in that particular experiment. We illustrate this procedure for evaluating feature 1's importance in an experiment where the lag=3 in Figure~\ref{fig:wif}.
\vspace{-2mm}
\item (2) We then calculate the feature's importance in each of the 91 experiments.  We then average the numbers to get the week-invariant feature importance weight.  
\end{description}

\vspace{-2mm}
Figures \ref{fig:randomized_logistic_regression_no_collab} to \ref{fig:randomized_logistic_regression_wiki_only} summarize these normalized average feature importance weights for different cohorts. 

\begin{figure*}[ht!]
   \centering
    \includegraphics[width=0.55\textwidth]{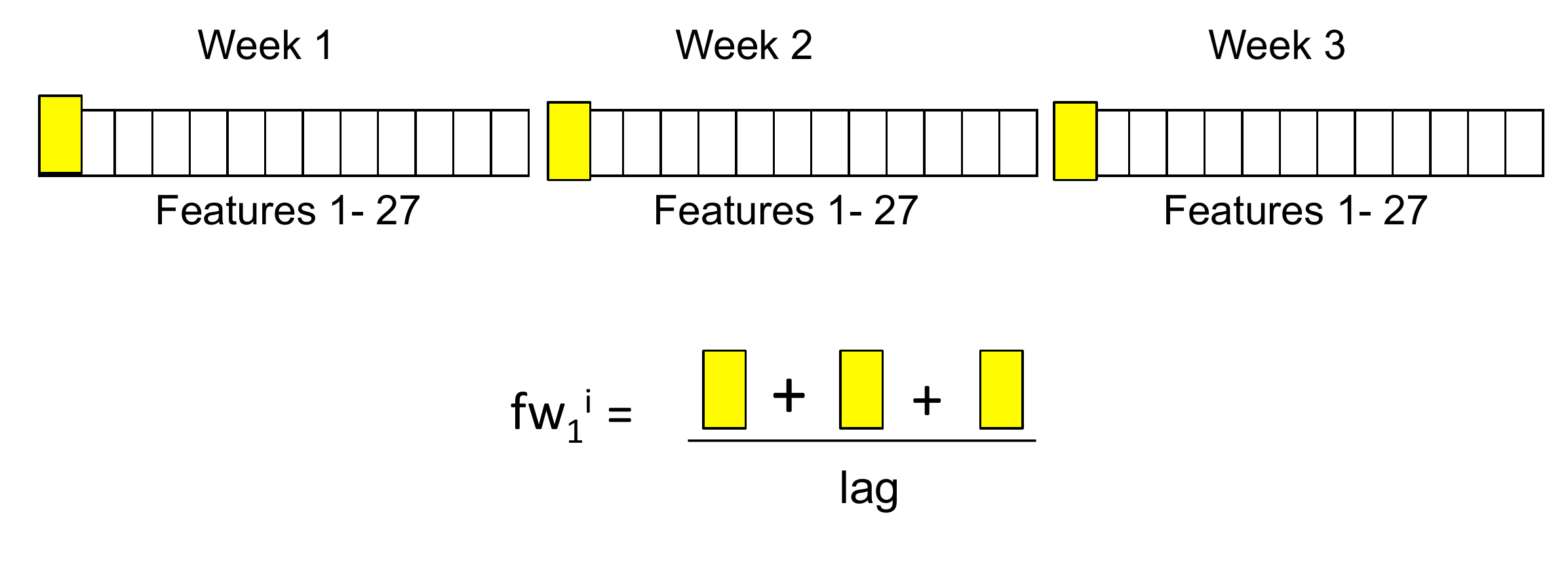}
     \caption{Aggregating feature 1's weights to assemble relative feature importance for a single experiment. In this example, the lag is 3. Three weeks data is used to predict a \sti in a future week. The Randomized logistic regression gives the weights for all 27 features for all three weeks (unnormalized). To assemble the week invariance relative weight for feature 1 we sum the weights and divide it with the total weights. We note that this is a heuristic. }\label{fig:wif}

\end{figure*}

\begin{figure*}[ht!]
  \centering
    \includegraphics[width=0.62\textwidth]{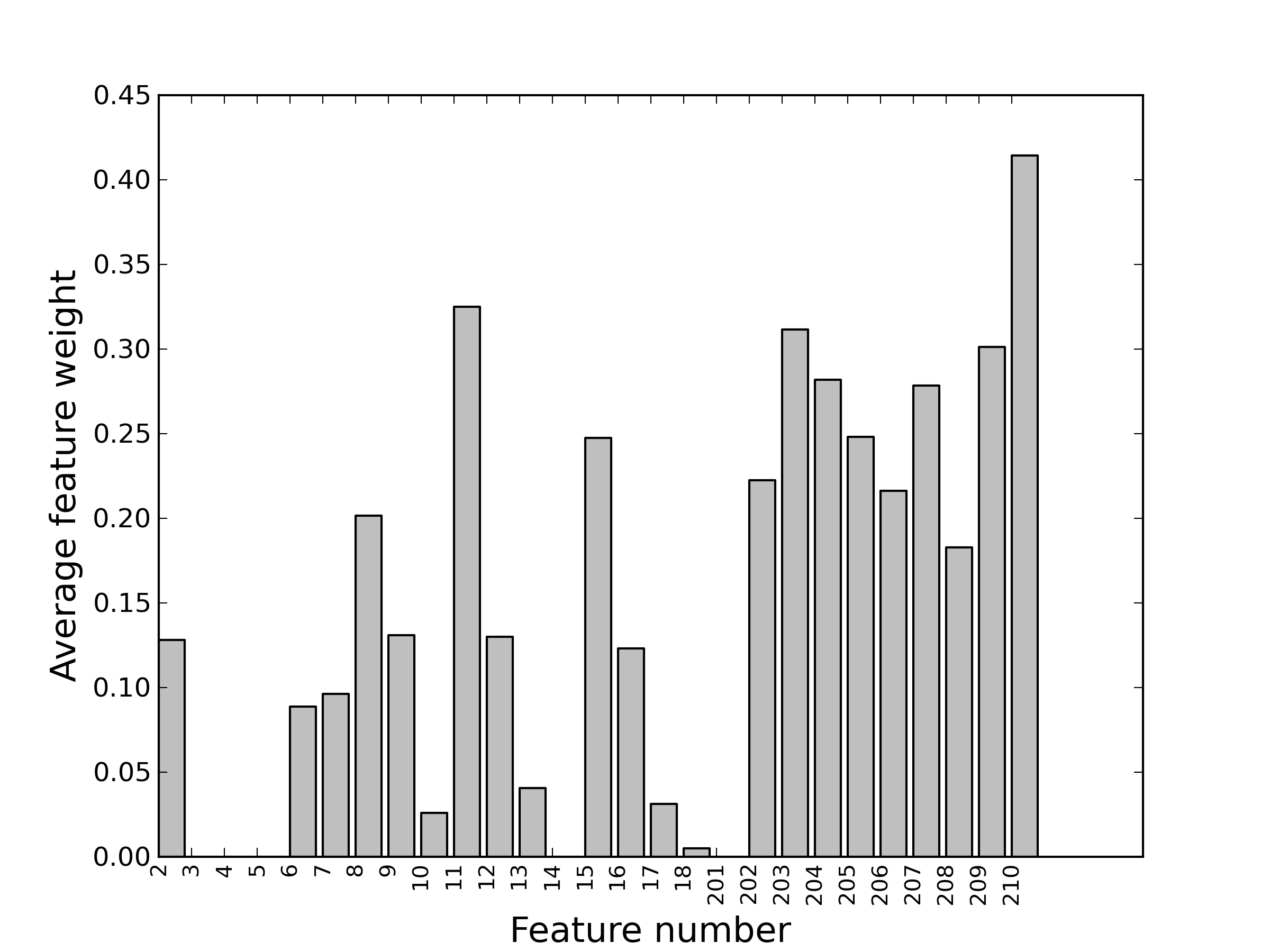}
     \caption{Feature importances for the \neither cohort. Top 5 features that had the most predictive power across multiple stopout predictive problems include \tten, \feleven, \tthree, \tnine, \tfive.}\label{fig:randomized_logistic_regression_no_collab}
\end{figure*}

The first thing that struck us as we looked at these plots was the difference in feature weights between the self-proposed features and the crowd-proposed features. In all four cohorts, the majority of the weight lies in the crowd-proposed features \ref{section:crowdself} (\x{201} through \x{210})! Clearly, the crowd can be utilized to a great degree. As features mostly represent complex and derived types, such as the percentiles (\x{202} and \x{203}), these plots suggest that those types of features have a very high predictive power. Additionally, they mostly involve the submissions table in MOOCdb. This includes the lab grade (\x{206}), pset grade (\x{207}) and predeadline submission time (\x{210})).

In the \neither cohort, the feature most indicative of \sti is the average predeadline submission time. The \forum cohort looks very similar, but uses a broader spectrum of features. In particular, we see that \x{5}, the average length of forum posts, is also highly predictive (of course, this could not have shown up in the \neither cohort, as by definition those students do not participate in the forum). Interestingly, we see a very low predictive power from the number of forum posts(\x{3}) and the number of forum replies (\x{201}), despite the fact that the length of the forum post is very important. This could imply that longer posts are indicative of more engagement in the course, or a greater mastery of the material.

\begin{figure*}[ht!]
  \centering
    \includegraphics[width=0.62\textwidth]{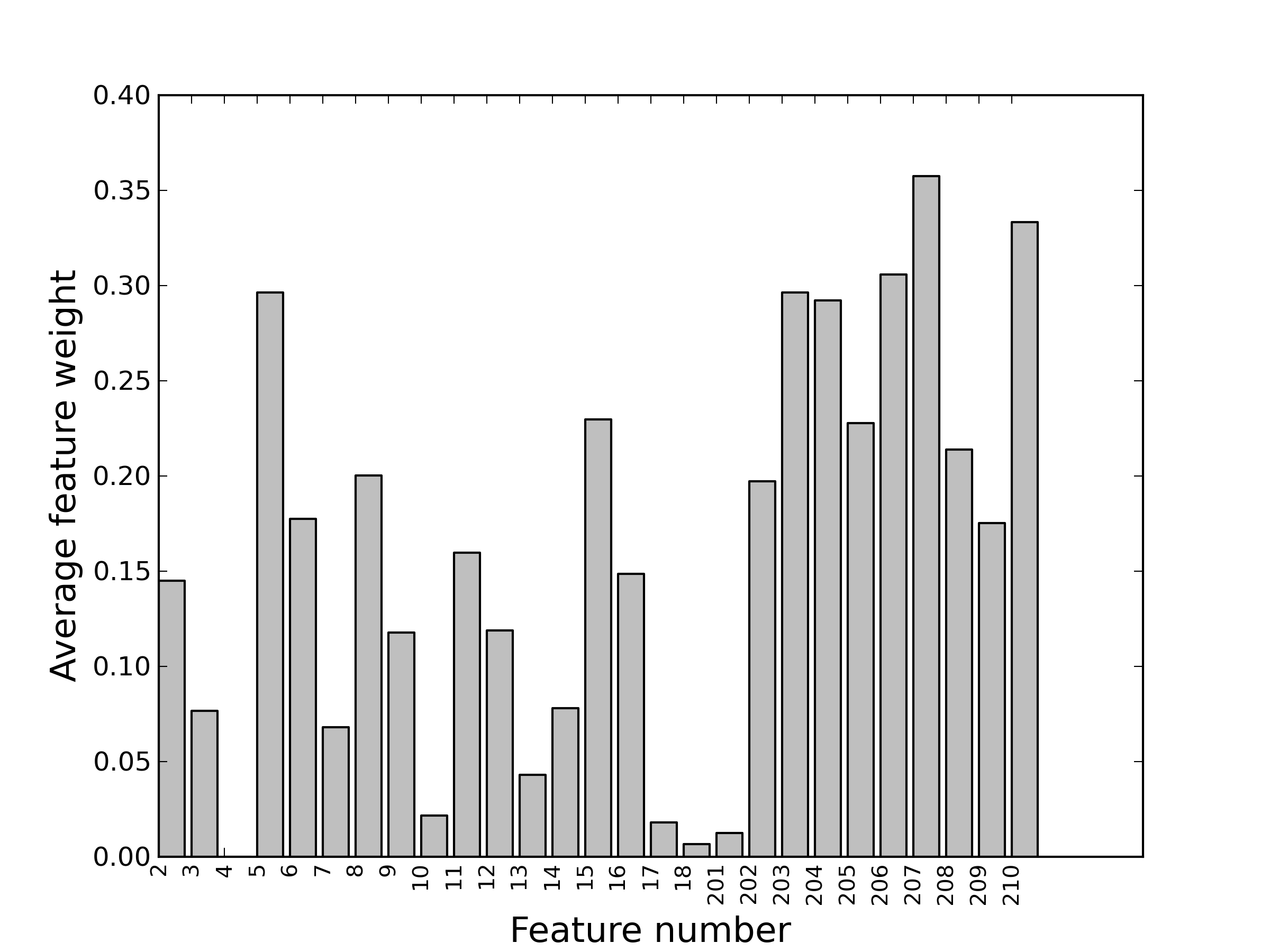}
     \caption{Feature importances for the \forum cohort. Top 5 features that had the most predictive power across multiple stopout predictive problems include \tseven, \tten, \ffive, \tsix, \tthree. }\label{fig:randomized_logistic_regression_forum_only}
\end{figure*}

In the both of our smaller cohorts, \both and \wiki, the lab grade (\x{206}) and lab grade over time (\x{207}) are the most predictive features. Although both of these cohorts participated in the Wiki, the number of Wiki edits (\x{4}) actually contains insignificantly small predictive power in both cases. Both cohorts show similar distributions overall. Similar to the larger cohorts, features related to submissions hold the most predictive power.

\begin{figure*}[ht!]
  \centering
    \includegraphics[width=0.62\textwidth]{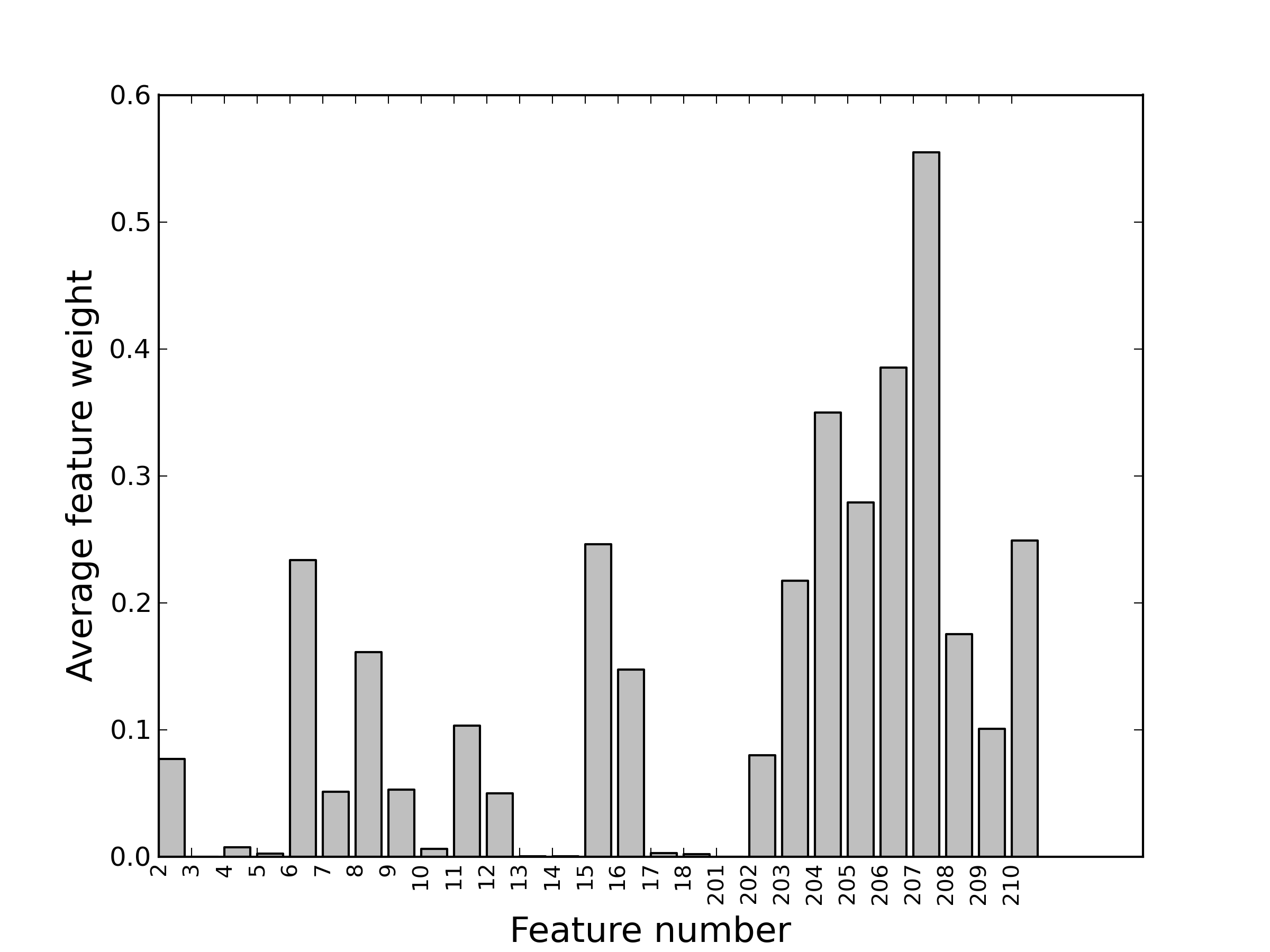}
     \caption{Feature importances for the \both cohort. Top 5 features that had the most predictive power across multiple stopout predictive problems include \tseven, \tsix, \tfour, \tfive, \tten. }\label{fig:randomized_logistic_regression_forum_and_wiki}
\end{figure*}

\begin{figure*}[ht!]
   \centering
    \includegraphics[width=0.62\textwidth]{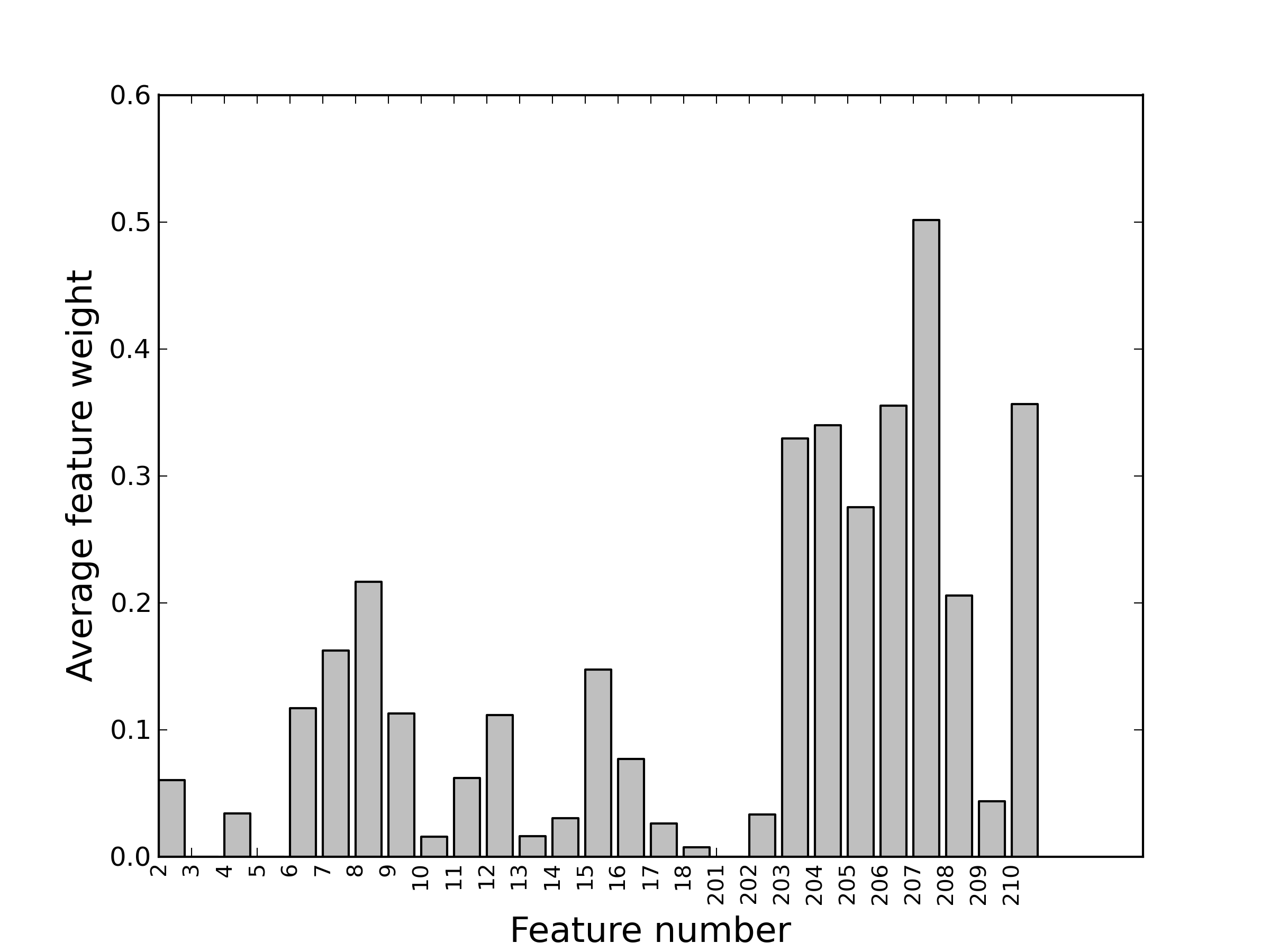}
     \caption{Feature importances for the \wiki cohort. Top 5 features that had the most predictive power across multiple stopout predictive problems include \tseven, \tsix, \tten, \tfour, \tthree. }\label{fig:randomized_logistic_regression_wiki_only}
\end{figure*}

\end{paragraph}

\section{Related work}\label{sect:related}
Efforts have been made by others to construct features that describe learner behavior in MOOCs longitudinally. Here in we present a few examples. \cite{kizilcec2013deconstructing} assemble a feature per-learner that has four categorical values during each assessment period. The four categorical values represent: on track (did assessment on time), behind (turned in assessment late), auditing (did not do assessment) and out (did not participate in the course at all). Computationally these can be captured by simply checking for submission activity in each assessment period. Authors note that they are easy to collect and are able to give powerful insights about learners engagement.  

\cite{halawadropout} extracted four features called \textit{video-skip}, \textit{assn-skip}, \textit{lag} and \textit{assn-performance}. The first two features inform whether a learner skipped videos in the previous week and skipped the assignments respectively. The third feature \textit{lag} checks if the learner is watching videos that are from previous weeks, if the learner is watching videos from week 2 and the course is currently in week 3 this feature's value is 1. The fourth feature measures learner's average quiz score. 

\cite{balakrishnan2013predicting} constructed 5 basic features, two of which, \sti and the number of forum posts (\x{1} and \x{3}), we independently used. The other three features are time spent on the lecture videos, number of threads viewed and number of times progress page was checked.   

\cite{yang2013turn} and \cite{ramesh:aaai14}, \cite{ramesh2013modeling} extract number of features from the forum activity of the learners. In \cite{yang2013turn} they are are length of post, thread starter (whether learner started threads or not) and content length (number of characters). Additionally, they construct a network of ``who talked to who" on the forums for every week and extract features on a per-learner basis from this network. \cite{ramesh:aaai14} extract counts for \textit{postActivity}, \textit{viewActivity} and \textit{voteActivity} from the learner interactions on the forums.  Additionally, they extract four binary values called \textit{posts}, \textit{votes}, \textit{upvote}, \textit{downvote} which are given a value 1 if the learner has engaged in that activity. They additionally tag the posts using an automated tool called \textit{OpinionFinder}. This tool tags each post as subjective/objective and positive/negative. The features are then the number of subjective posts the learner made and the number of the positive posts the learner made out of the total number of posts. 

One of the limitations of the last three efforts is that they focus primarily on forum activity. While we extract some of these features as well we note that many learners in fact do not participate in forums while still actively engage in the course. For example in the course we consider in this paper, only 7860 students participate in the forums out of a total of 105622.  Hence, the analysis via these features is limited to only a small proportion of the learners. 

We note that, as per our categorization of features in Section~\ref{sect:challenges} most of these features fall into the \textit{simple} category. Many of these features solely access one mode of student activity, for example submissions/assessments, and do not require combining additional information like correctness of the problem submissions. Many of these features we independently extracted in \selfself and have evaluated their predictive power. For example, amount of the time spent on lecture videos did not appear to have significant predictive power when compared to other complex features.  

Our extraction effort is the first instance, to our knowledge, wherein an extensive, sophisticated feature set has been constructed on MOOC behavioral data. We continue to add to this set and are accumulating a massive set of feature ideas and scripts that will seamlessly run on the data model and are available for sharing and re-use.


\section{Research findings and conclusions}\label{sect:conclusions}

\noindent \textbf{Finding 1: Features proposed by crowd for the \sti prediction problem mattered}: For all four cohorts we found that features proposed by the crowd mattered significantly more than the features we \selfself. 

\noindent \textbf{Finding 2: Different features mattered for different cohorts}: We also found interesting differences between features that mattered between different cohorts. For example, for the \neither cohort features that explain students success in assignments mattered along with the average pre deadline submission time. For cohorts that consisted of students that interacted with other students, lab grade over time mattered consistently. For cohort that participated on forums only, the length of the forum post is a good indicator whether they are likely to stopout. 

\noindent \textbf{Finding 3: Complex and derived features mattered more} We also found the more influential features were quite nuanced and complex. They incorporated data from multiple modes of learner activity (submissions, browsing and collaborations), required carefully linking data fields.  Relational features that compared a learner to others and statistical summaries were proposed by the crowd and mattered quite a bit. 


\noindent \textbf{What is next?} 

\noindent \textbf{Addressing the challenges of feature discovery}: Human intuition and insight defy complete automation and are integral part of the process. We conclude that the best way to address this challenge is to involve as many people (experts, instructors, students, researchers as possible). People can play multiple roles. They can propose ideas or concepts from which variables can be formed, help us extract variables given mock data, or validate many ideas that we might ourselves have. To enable this, our next goal is to scale this process. We are currently designing a web based platform for collaborative feature definition and discovery. 

\noindent \textbf{Addressing the challenges of curation and processing}  To address these challenges, we propose :

\begin{itemize}
\item We propose the sharing and reuse of feature engineerings scripts, such as those used in this paper. that we be able widely share and re-use feature generation scripts we used in this paper. We have made sharing possible by standardizing the data schema so all our scripts can be used for multiple courses. We are currently testing our scripts across courses to assess their reuse value on different platforms \footnote{To date edX and Coursera}. 
\item We anticipate, given inherent differences among courses, that some features will be present, others will need adjustment and still new ones may yet be engineered and their scripts shared. This is an area in which we are currently active. We endorse a promising direction towards  shared methods that operationalize longitudinally \textit{per-learner} variables across MOOCs.
\end{itemize}

\bibliography{mooc} 
\bibliographystyle{icml2013}

\end{document}